\documentstyle[epsf]{l-aa}     

\begin{document}

   \thesaurus{4          
	      (11.07.1;  
	       11.19.2;  
	       11.16.1;  
	       13.09.1)  
	     }
   \title{Near-infrared surface photometry of spiral galaxies
   \thanks{Based in part on observations collected at the European Southern
Observatory, La Silla, Chile.}
   \fnmsep
   \thanks{Tables 2a and 2b are presented in electronic form only; Tables
1 through 3 are available from the CDS, Strasbourg 
(anonymous ftp to 130.79.128.5).}
    }
   \subtitle{I. The data}

   \author{Ph. H\'eraudeau \inst{1} \and F. Simien \inst{1} 
\and G. A. Mamon \inst{2,3} }


   \institute{Observatoire de Lyon, F-69561 Saint-Genis-Laval cedex, France 
\and Institut d'Astrophysique (CNRS), 98 bis boulevard Arago, F-75014 Paris,
France
\and DAEC, Observatoire de Paris, F-92195 Meudon, France}

   \date{Received ..................; accepted 22 Dec 1995}

   \maketitle

   \markboth{Ph. H\'eraudeau et al.: Near-infrared surface photometry of spiral
galaxies}{}

   \begin{abstract}
We present $K'$-band surface photometry of a sample of 31 inclined 
Sa-Scd galaxies, together with additional $J$- and $H$-band data for four of 
them. In this first paper of a series, profiles are presented, together with
global and isophotal parameters.
Our profiles are compared to similar $B$, $R$ and $I$ data collected from 
other sources. 
Three galaxies exhibit previously unknown small bars in their center, while
in five others, such bars may also be present.
Four objects present a narrow elongated feature in their center aligned with
their major axis, which could be an inward extension of the disk.
A few galaxies display very thin spiral arms.
Color-color diagrams indicate that the extinction inside the four galaxies for
which we have $JHK'$ images is limited to $A_V < 2$.

   \end{abstract}
   \section{Introduction}
The recent availability of near-infrared (NIR) arrays and their continuous
improvement have 
resulted in a fast growing amount of published surface-photometry data on 
spiral galaxies (de Jong \& van der Kruit 1994; Peletier et al. 1994; 
Terndrup et al. 1994; McLeod \& Rieke 1995; Rauscher 1995, to cite only a 
few recent works dealing with sizable samples).
This paper is the first in a series dedicated to the study of several 
structural and kinematical aspects of spirals in a wide range of Hubble 
types; it is mainly limited to the reduction of the surface photometry and
to the discussion of morphological aspects. It is organized as follows: sample
selection, observational procedures and preliminary reductions are presented
in Sect. 2, with complementary data in other bandpasses (collected from the
literature) described in Sect. 3. Section 4 compares our photometry
with that of other sources, sums up the main morphological properties, and
discusses 
briefly color-color diagrams for the four objects observed in three passbands;
the last subsection is a discussion which presents the planned applications,
mainly the decomposition into the main stellar components and the derivation 
of mass models.
   \section{Observations and data reductions}
   \subsection{Sample}
As part of a wider program of surface photometry observations on nearby
galaxies, we have selected an initial sample of southern, unbarred 
galaxies from the following criteria: (i) a morphological type
from Sa to Scd, (ii) an absolute magnitude $M_{\rm B}<-19$ and a distance
$\Delta < 53.3$ Mpc (with $H_0={\rm 75 \, km \, s^{-1} \, Mpc^{-1}}$,
this corresponds to a radial velocity $cz<4000 \, {\rm km \, s^{-1}}$, after
a correction of 220 km s$^{-1}$ for the Virgocentric infall, from Tammann \&
Sandage 1985), (iii) an inclination angle $45\degr < i < 78\degr$ (based on an 
outer axial ratio $0.21 < R_{25} < 0.71$), (iv) an apparent diameter 
$D_{25}<5.5'$, and (v) coordinates restricted to $-72\degr< \delta < 0\degr$,
with a zone of avoidance at $|b_{\rm II}|<20\degr$.

This first selection brought out 168 galaxies. Telescope 
availability for a single run cut down to 93 the number of observable
objects, still 
representing a volume-limited sample; 31 objects were actually observed,
among which 22 strictly meet the criteria above, while nine other
galaxies were added for their specific interest, although they were up to
17\% fainter than the limit $M_B = -19$ and/or at moderate northern
declination. 
Histograms of the main
characteristic parameters showed that the distribution within the observed
set does not differ fundamentally from that in the selected sample, 
except for a deficiency of far-away objects; in contrast, our set includes more
small-$D_{25}$ objects than the sample limited by 
$cz<3200 \, {\rm km \, s^{-1}}$. To sum up, our observed set of 31 galaxies
constitutes only marginally a statistically homogeneous sample.

   \subsection{Observations}
We used the IRAC2-A camera equipped with a NICMOS3 $256 \times 256$ detector, 
attached to
the ESO/MPI 2.2-m telescope at La Silla, Chile, between July 5 and 9, 1993.
This instrument has been extensively tested and described (for details, see,
e.g., Moorwood et al. 1992). All 31 galaxies were observed in the $K'$ filter
(2.1 $\mu\rm m$), located slightly shortward of the standard $K$ passband
in order to reduce the thermal background (Wainscoat \& Cowie 1992); four of
these objects, selected for their appropriate dimensions and orientation, were,
in addition, observed in the $J$ ($1.25 \mu\rm m$) and
$H$ ($1.65 \mu\rm m$) bands, with the aim of getting some insight in dust
obscuration, through two-color indices.

For the camera magnification, we selected objective C, which appeared to 
be the best suited to our program, in terms of compromise
between resolution, field, and flatness of response: with this setting, the
field was $125 \times 125$\arcsec, and the pixel size 0.49\arcsec.

For good sky subtraction, several galaxy (G) and sky (S) exposures were made 
for each object, in S-G-S sequences. The different galaxy frames were shifted 
in position by a few 
arcsec for a better correction of bad pixels on the resulting, average
frame; a few large objects are actually a mosaic of frames separated by a
fraction of an arcmin. Typical individual exposure times on a galaxy ranged 
from 5 to 12 seconds, for a total of $\approx$ 10--20 minutes: the cumulated
on-object
exposure times in $K'$ are listed in Table 1. For the four galaxies also
observed in $J$ and $H$, they were 30--40\% shorter than the $K'$ time (this
difference was motivated by the high luminosity of these objects); these 
shorter integrations do not reflect into statistically larger error bars 
in the profiles of Fig. 4: these errors are even often smaller (especially in
$J$), due to the fainter sky brightness with respect to $K'$, but we note 
a lesser typical quality for the $H$-band data. The successive sky frames were 
separated by less than five minutes in time, and several locations were 
selected around the galaxy, at a few arcmin from its center. The integration 
time on sky was either the same as on the object, or half that amount.

During our four-night run, about one night was lost due to poor weather. On
another night, some thin cirrus was present close to a few objects, making
flatfielding difficult and calibration uncertain (details are given in
Appendix  
A: comments on individual objects). The observations of standard stars were
also used to monitor the atmospheric transparency (see Sect. 2.3). On the 
whole, the seeing conditions, as measured by a two-gaussian fit to star 
images, ranged from 1.0\arcsec \ to 2.0\arcsec \ (FWHM).
   \subsection{Preliminary reductions}
   \begin{enumerate}
   \item Flatfielding: we adopted as the standard flatfield the difference 
between frames on the dome obtained, successively, with and without 
illumination by a tungsten source; this procedure allowed to remove the large 
dark ring present on every single frame of IRAC2-A, and, more generally, all
``dark current'' manifestations. Several such exposures 
were made at the beginning and at the end of each night. A couple of bands
with a slightly different mean brightness were also present on the raw frames,
and their contrast turned out to be sometimes variable; so, when processing
images for which this was noticed, the dome flatfield was replaced
by a neighboring sky frame, close to the galaxy frame in both time and
position. 
At this stage, we noticed $\approx$ 0.5\% of bad pixels on the detector.
   \item Sky brightness determination: in the standard procedure, i.e., when
using the dome flatfields, and after dividing all frames by these,
we subtracted from the galaxy a mean of neighboring sky frames. 
On the resulting, average galaxy frame, we measured the background in several
outer areas, and from its dispersion we estimated the degree of flatness: 
it was, typically, better than 0.07\% of the sky level.
Figure 1 presents the variations of the mean sky level in the
$K'$ band, from the beginning to the end of our run; this monitoring was
made by collecting all the average sky brightnesses for the observed galaxies,
corrected to the same airmass ($=1.0$). For each night, it is worth noting that
the trend toward a darker level at the middle of the night is $\simeq 0.5$
mag arcsec$^{-2}$; the sky color indices are $J-H \simeq 2.0$ 
and $H-K' \simeq 0.4$.
   \item Photometric calibration: we used faint stars from the list of 
Carter (1993), and our magnitudes are thus consistent with the SAAO system; 
during the three clear nights, at airmasses between 1.0 and 2.0, we observed 
either four or five stars, each one with its image successively in five 
different locations on the detector. The calculated extinction coefficients 
have mean values of 0.14 in $J$, 0.10 in $H$, and 0.11 in $K'$, not
inconsistent with other observations at La Silla (Engels et al. 1981;
Bersanelli et al. 1991). With these data, we estimate the uncertainty of our 
calibrations to be $\simeq 0.05$ mag. For the
calibration stars, photometric corrections from $K$ to
$K'$ were made according to Wainscoat \& Cowie (1992).
   \end{enumerate}
   \subsection{Isophote analysis}
On the calibrated images, stars were removed, if any, and the ellipse-fitting
program of the ESO-Midas reduction software was run, resulting in values
of the following parameters: brightness level $\mu$, ellipticity $\epsilon$ 
($\equiv 1 -$ axial ratio), and position angle $PA$ at a range of
semi-major axis $r$. 
The values of $PA$ and $\epsilon$ listed in Table 3 correspond to 
$\mu_{K'} = 19.5$. At faint levels, $\mu$ was determined by integrating 
within elliptical annuli of increasing ``thickness''; for the outermost 
isophotes, fixed ellipticity and orientation were assumed, so as to lower 
the uncertainty on $\mu$. 

In addition, we analyzed the departure of isophotal
shapes from perfect ellipses, in the inner region where this is relevant to
the  
detection of bars, and disky, boxy, or triaxial structures. We adopted the
technique of Michard \& Simien (1988), and we parameterized an isophote shape
(assumed to be symmetric) by the Fourier coefficients $e_4$, $e_6$, and $e_8$. 
Although not strictly identical to the well-known coefficient 
$a_4$, our $e_4$ describes the isophotes in the same way, with positive and 
negative values referring, respectively, to ``disky'' and ``boxy'' shapes. 
Needless to say, this analysis requires much better data than the simple
ellipse fitting, and we limited ourselves to: a) galaxies for which this 
was possible well outside the central region dominated by resolution effects,
and b) regions where the isophotes appeared reasonably symmetric, free from
chaotic distortions. The analysis has been possible on 12 of our sample 
galaxies. Figure 4 presents the $e_4$ profiles, together with the $e_6$
for the four large galaxies observed in the three passbands; Table 2b lists
$e_4$ through $e_8$ ($e_4$ alone is sometimes insufficient for the 
reconstruction of the isophote, higher-order terms are then used, usually
up to $e_6$ or $e_8$). 
   \subsection{Determination of integrated luminosities}
The total luminosity of each galaxy is computed by extrapolating exponential
models, adapted to the measured outer profiles, 
beyond the range of reliable isophote analyses,
using fixed isophote ellipticities and orientations. Tests show that
the uncertainties on the integrated luminosities are $\simeq 0.1$ mag. 
   \subsection{Comments on individual galaxies}
Miscellaneous remarks on either the morphology, the photometric
characteristics, 
or peculiarities appeared during the reduction procedure are presented in 
Appendix A.

   \begin{figure}
   \epsfxsize=8.8cm
   \epsffile[59 243 479 448]{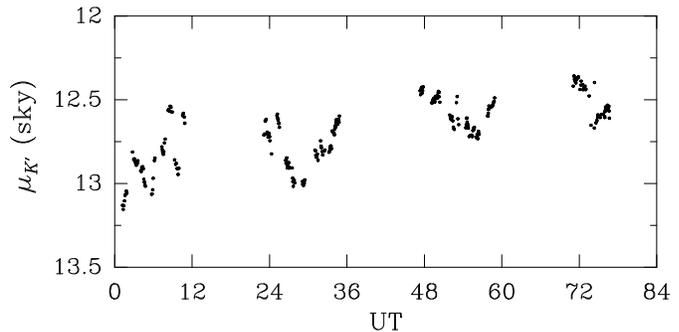}
   \caption{Variation of the $K'$-band brightness of the night sky during the
four-night run. Abscissae give UT starting on July 5, 1993}
   \end{figure}
%
    \begin{table*}
    \caption{Catalog elements and exposure times for observed galaxies}
    \begin{flushleft}
        \begin{tabular}{rllcccccccr}
        \hline
\# & \multicolumn{2}{c}{object}& $T$ & $\alpha_{1950}$ & $\delta_{1950}$ &
$B_{\rm T}$ & 
 $D_{25}$ & $cz$ & $\Delta$ & exp\\
        \hline
(1) & \hspace{0.4cm} (2) & \hspace{0.3cm}(3) & (4) & (5) & (6) & (7) & (8) &
(9) & (10) & (11) \\ 
        \hline
 1 & E 320-26    &                 &   3 &  11 47 19.0  & $-38$ 30 24  &
12.84 &  2.4 &  2835 &     35.7 &  768 \\   
 2 & E 380-06    &                 &   3 &  12 12 57.0  & $-35$ 21 06  &
12.66 &  3.7 &  2943 &     37.4 &  720 \\   
 3 & E 267-37    &  N 4219         &   4 &  12 13 50.0  & $-43$ 02 48  &
12.71 &  4.2 &  1986 &     24.2 &  800 \\   
 4 & E 322-28    &                 &   2 &  12 32 04.0  & $-42$ 15 24  &
14.41 &  1.4 &  3715 &     47.4 & 1200 \\   
 5 & E 323-38    &                 &   6 &  12 51 30.4  & $-41$ 33 04  &
14.31 &  1.3 &  3339 &     42.5 & 1200 \\   
 6 &             &  N 5119         &   5 &  13 21 21.5  & $-12$ 00 58  &
13.07 &  1.3 &  2857 &     38.1 & 1200 \\   
 7 &             &  N 5604         &   1 &  14 22 06.9  & $-02$ 59 12  &
13.54 &  1.8 &  2747 &     37.4 &  900 \\   
 8 &             & M -2-38-22      &   3 &  14 49 28.1  & $-10$ 32 11  &
14.01 &  1.6 &  2587 &     34.8 & 1200 \\   
 9 &             & M -1-39-05      &   2 &  15 20 42.8  & $-03$ 58 30  &
13.65 &  1.6 &  1999 &     27.5 & 1200 \\   
10 &             & N 5937          &   3 &  15 28 09.9  & $-02$ 39 35  &
12.71 &  1.8 &  2805 &     38.4 & 1200 \\   
11 &             & N 6063          &   6 &  16 04 48.1  & $+08$ 06 40  &
13.67 &  1.7 &  2847 &     39.7 & 1200 \\   
12 & E 139-11    &                 &   2 &  17 31 52.0  & $-58$ 13 24  &
14.22 &  1.5 &  2736 &     33.9 &  600 \\   
13 & E 140-24    & I 4717          &   3 &  18 28 58.0  & $-58$ 00 42  &
14.16 &  1.5 &  3168 &     39.6 &  960 \\   
14 & E 184-17    & I 4821          &   5 &  19 05 27.0  & $-55$ 05 48  &
13.65 &  1.8 &  2698 &     33.4 & 1200 \\   
15 & E 184-27    & I 4826          &   2 &  19 08 08.0  & $-57$ 17 06  &
14.47 &  1.4 &  3459 &     43.5 & 1200 \\ 
16 & E 184-67    & N 6788          &   2 &  19 22 47.0  & $-55$ 03 06  &
12.87 &  2.9 &  3168 &     39.7 &  800 \\   
17 & E 142-35    & N 6810          &   2 &  19 39 21.0  & $-58$ 46 30  &
12.35 &  3.0 &  1941 &     23.1 &  800 \\   
18 & E 073-22    & I 4929          &   3 &  20 01 14.0  & $-71$ 49 30  &
14.33 &  1.4 &  4194 &     52.6 & 1200 \\   
19 & E 233-41    & N 6870          &   2 &  20 06 33.0  & $-48$ 26 06  &
13.18 &  2.6 &  2620 &     32.6 &  750 \\   
20 & E 399-25    &                 &   1 &  20 10 11.0  & $-37$ 20 24  &
13.78 &  2.2 &  2540 &     32.2 & 1200 \\   
21 & E 287-13    &                 &   4 &  21 19 56.0  & $-45$ 59 12  &
13.01 &  2.9 &  2695 &     33.5 &  960 \\   
22 & E 107-36    & N 7083          &   5 &  21 31 50.0  & $-64$ 07 42  &
11.90 &  3.6 &  3111 &     38.2 &  840 \\   
23 & E 466-38    & N 7172          &   1 &  21 59 07.0  & $-32$ 06 36  &
12.80 &  2.4 &  2566 &     32.4 &  900 \\   
24 &             & N 7328          &   2 &  22 34 59.8  & $+10$ 16 23  &
13.96 &  2.1 &  2824 &     38.4 & 1200 \\   
25 & E 290-29    & I 5267          &   1 &  22 54 22.0  & $-43$ 39 48  &
11.31 &  5.4 &  1714 &     20.2 &  600 \\   
26 & E 406-34    & I 5271          &   3 &  22 55 16.0  & $-34$ 00 36  &
12.47 &  2.6 &  1720 &     20.8 & 1200 \\   
27 &             & N 7537          &   4 &  23 12 01.9  & $+04$ 13 33  &
13.86 &  2.1 &  2675 &     35.8 &  900 \\   
28 &             & N 7541          &   4 &  23 12 10.3  & $+04$ 15 43  &
12.42 &  3.3 &  2682 &     35.9 &  600 \\   
29 &             & N 7606          &   3 &  23 16 29.2  & $-08$ 45 33  &
11.51 &  4.8 &  2232 &     29.0 & 1200 \\   
30 & E 605-07    & I 5321          &   1 &  23 23 43.0  & $-18$ 13 48  &
13.96 &  1.2 &  2867 &     36.8 &  720 \\   
31 &             & N 7721          &   5 &  23 36 14.1  & $-06$ 47 43  &
12.22 &  3.0 &  2013 &     26.1 &  600 \\   
        \hline
        \end{tabular}
    \end{flushleft}
{\it Columns\/}:

(1), (2), (3)  galaxy identification\\
(4) $T$: morphological type\\
(5), (6) coordinates\\
(7) $B_{\rm T}$: integrated blue magnitude\\
(8) $D_{25}$: apparent diameter at brightness level $\mu_B=25$,
in arcmin\\
(9) $cz$: heliocentric radial velocity in ${\rm km \, s^{-1}}$\\
(10) $\Delta$: distance in Mpc (with $H_0={\rm 75 \, km \, s^{-1} \, 
Mpc^{-1}}$), from $cz$ corrected for Virgocentric infall\\
(11) exp: total $K'$ exposure time, in seconds \\
{\it Note\/}: columns (4) to (9) are from the {\it LEDA\/} database
    \end{table*}
%
   \section{Additional surface-photometry data}
With the aim of making relevant comparisons, we also collected available
surface-photometry data in other passbands for our sample galaxies. The sources
are the following:
\begin{enumerate}
\item $B$- and $R$-band images were requested from the ESO-LV database
(Lauberts \& Valentijn 1989); for these, the pixel size is 1.35\arcsec\ and the
seeing conditions were from 2\arcsec \ to 3\arcsec \ (FWHM). The images 
were available for 20 of our galaxies; they come from scans on Schmidt
plates and, expectedly, many of them lack the high S/N ratio of recent CCD 
data, especially the $B$ frames, but they are still valuable for bringing out
extinction effects. Since offsets have been reported in the photometric
calibration of some of these images (Huizinga 1994; Peletier et al. 1994),
we made careful checks; indeed, for 14 out of 20, we noticed a discrepancy 
between the sky brightnesses indicated in the headers of the images, and the 
values that we actually measured within selected areas. We found a mean 
magnitude difference of $0.23 \pm 0.01$ in $B$ and $0.13 \pm 0.02$ in $R$ 
(the sky measured on the images being fainter), thus confirming values given 
by the above-cited authors. For nine galaxies, aperture-photometry from the 
catalogs of Longo \& de Vaucouleurs (1983) and de Vaucouleurs \& Longo (1988)
allowed a direct re-calibration; 
for the others, we have assumed that the sky brightness of the header was 
correct (we note that Peletier et al. assumed instead that the value given 
in the ESO-LV catalog was correct), and we have re-calibrated accordingly.
After these corrections, the ellipse-fitting program was run, resulting in
the adopted $B$ and $R$ profiles.
\item For 10 galaxies, $I$-band data were obtained from Mathewson et al. 
(1992). These authors indicate that most of their galaxies with $D_{25} 
\la 5'$ were observed with a pixel size of 0.56\arcsec, and
they do not specify their typical seeing dimension. They determined the
photometric profiles from ellipse fitting to the isophotes; their  magnitudes,
in the Kron-Cousins system, are estimated accurate to within 0.05 mag. 
\item Data from the Haute-Provence Observatory: in their visible- and
$I$-band surface photometry, H\'eraudeau \& Simien (1996: hereafter HS96) have
four objects in common with the present sample; their data are in the 
Kron-Cousins system and have been obtained with a pixel size of 0.78\arcsec.
For NGC 7541 their $B$ image, and $V$ and $I$ profiles (seeing conditions of 
2.5\arcsec) are presented in our Fig. 4; for NGC 7606, NGC 7537, and NGC 7721,
the $V$ profiles are shown in Fig. 4, but due to poor seeing conditions 
($\simeq 3.5''$), they are unreliable for $r \la 5''$. 
\end{enumerate}

    \addtocounter{table}{1}
%
    \begin{table*}
    \caption{Results of the photometric analysis}
    \begin{flushleft}
    \begin{tabular}{llrrrrrrrrrrr}
    \hline
\# & \multicolumn{2}{c}{object} & $PA$ & $\epsilon$ & $J_{\rm T}$ & 
$H_{\rm T}$ & $K_{\rm T}$ & $D_{25}(B)$ & $D_{20}(J)$ & $D_{19}(H)$ & 
$D_{19}(K)$ & $D_{20}(K)$ \\
    \hline
(1) & \hspace{0.4cm}(2) & \hspace{0.3cm}(3) & (4) & (5) & (6) & (7) & (8) &
(9) & (10) & (11) & (12) & (13) \\
    \hline
 1 & E 320-26    &                 & 162 & 0.63 & 9.78 & 9.02 & 8.71 & 153.2
&  97.2 &  87.4 & 100.0 &  .... \\ 
 2 & E 380-06    &                 &  71 & 0.55 &      &      & 7.74 & 213.7
&       &       & 127.6 &  .... \\ 
 3 & E 267-37    &  N 4219         &  30 & 0.65 & 9.20 & 8.36 & 8.10 & 197.7
& 134.4 & 127.7 & 133.3 &  .... \\ 
 4 & E 322-28    &                 & 111 & 0.58 &      &      &10.36 &  86.0
&       &       &  37.1 &  61.0 \\ 
 5 & E 323-38    &                 & 178 & 0.52 &      &      &10.29 &  80.2
&       &       &  46.5 &  58.3 \\ 
 6 &             &  N 5119         &  22 & 0.69 &      &      & 9.82 &
.... &       &       &  58.6 &  69.2 \\ 
 7 &             &  N 5604         &  10 & 0.41 &      &      & 9.71 &
.... &       &       &  45.0 &  62.8 \\ 
 8 &             & M -2-38-22      & 104 & 0.43 &      &      &10.03 &
.... &       &       &  33.8 &  53.3 \\ 
 9 &             & M -1-39-05      &  26 & 0.42 &      &      &10.60 &
.... &       &       &  41.0 &  58.7 \\ 
10 &             & N 5937          &  25 & 0.61 &      &      & 9.12 &
.... &       &       &  58.8 &  73.8 \\ 
11 &             & N 6063          & 168 & 0.34 &      &      & .... &
.... &       &       &  24.4 &  43.2 \\ 
12 & E 139-11    &                 &  88 & 0.67 &      &      &10.05 &  91.5
&       &       &  43.1 &  53.7 \\ 
13 & E 140-24    & I 4717          &  93 & 0.75 &      &      & 9.42 & 102.9
&       &       &  72.8 &  86.7 \\ 
14 & E 184-17    & I 4821          &   5 & 0.58 &      &      &10.24 & 117.0
&       &       &  55.5 &  79.5 \\ 
15 & E 184-27    & I 4826          &  41 & 0.34 &      &      &10.86 &  78.3
&       &       &  27.2 &  40.6 \\ 
16 & E 184-67    & N 6788          &  71 & 0.66 & 9.82 & 8.84 & 8.53 & 175.3
&  98.2 &  99.6 & 105.8 &  .... \\ 
17 & E 142-35    & N 6810          & 174 & 0.73 & 8.98 & 8.15 & 7.81 & 193.6
& 130.6 & 123.8 & 125.0 &  .... \\ 
18 & E 073-22    & I 4929          &  20 & 0.75 &      &      &11.09 & 111.3
&       &       &  49.4 &  58.0 \\ 
19 & E 233-41    & N 6870          &  87 & 0.54 &      &      &10.19 & 151.9
&       &       &  46.1 &  72.4 \\ 
20 & E 399-25    &                 & 167 & 0.41 &      &      & .... &  97.0
&       &       &  30.0 &  47.4 \\ 
21 & E 287-13    &                 &  61 & 0.61 &      &      & 9.28 & 229.0
&       &       &  93.4 & 123.9 \\ 
22 & E 107-36    & N 7083          &  16 & 0.45 &      &      & 8.23 & 224.4
&       &       &  90.9 & 130.8 \\ 
23 & E 466-38    & N 7172          & 104 & 0.35 &      &      & 8.27 &
.... &       &       &  91.7 & 116.7 \\ 
24 &             & N 7328          &  88 & 0.65 &      &      & 9.50 &
.... &       &       &  60.4 &  93.5 \\ 
25 & E 290-29    & I 5267          & 139 & 0.25 &      &      & 7.81 & 347.7
&       &       &  94.1 &  .... \\ 
26 & E 406-34    & I 5271          & 139 & 0.62 &      &      & 8.22 & 162.6
&       &       & 113.9 & 142.3 \\ 
27 &             & N 7537          &  78 & 0.72 &      &      &10.00 &
.... &       &       &  47.7 &  74.8 \\ 
28 &             & N 7541          &  98 & 0.73 &      &      & 8.44 & 198.3
&       &       & 131.7 &  .... \\ 
29 &             & N 7606          & 149 & 0.56 &      &      & 8.01 &
.... &       &       & 135.2 & 188.0 \\ 
30 & E 605-07    & I 5321          &  50 & 0.40 &      &      &11.35 &  76.3
&       &       &  16.3 &  35.6 \\ 
31 &             & N 7721          &  16 & 0.64 &      &      & 8.89 &
.... &       &       & 100.6 & 151.1 \\ 
    \hline
    \end{tabular}
    \end{flushleft}
{\it Columns\/}: \\
(1), (2), (3): galaxy identification\\
(4) $PA$: position angle of outer isophotes (at $\mu_{K'}=19.5$), in degrees,
North through East \\
(5) $\epsilon$: outer ellipticity (at $\mu_{K'}=19.5$) \\
(6), (7), (8) $J_T$, $H_T$, $K_T$: total magnitudes in the $J$, $H$, and $K$ 
(from $K'$ data) bands\\
(9) $D_{25}(B)$: major axis in arcsec at $\mu_{B} = 25$, recomputed on ESO-LV
images with a corrected sky brightness, when relevant
(10), (11), (12), (13) $D_{20}(J)$, $D_{19}(H)$, $D_{19}(K)$, $D_{20}(K)$: 
major axis in arcsec at $\mu_{J} = 20$, $\mu_{H} = 19$, $\mu_{K} = 19$,
and $\mu_{K} = 20$ (from $K'$ data) \\
    \end{table*}
%
   \section{Global results and discussion}
   \subsection{Presentation of the results}
Greyscale images and isophotes of the galaxies are presented in $K'$ and, 
when available, in $B$, $J$, and $H$, in Fig. 4 (Table 5 in Appendix B 
gives the lowest-level isophote for each object). Profiles are given for 
surface brightness (in one or more passbands: $B$, $R$, $I$, $J$, $H$, and/or
$K'$), colors (when available), ellipticity, and position angle; profiles are
also given for 
isophotal-shape coefficient $e_4$ (when it was possible to calculate its
value at more than a couple of arcsec), and, for the four galaxies with 
$JHK'$ observations, $e_6$ ({\it ibid\/}).

Table 2a collects the luminosity profiles (for convenience, we have added to
our NIR data the $B$, $R$, and/or $I$ data), the ellipticities and orientations;
Table 2b lists the $e_4$ through $e_8$ coefficients. These two Tables are
proposed in electronic form only.
 
Table 3 collects total magnitudes, parameters of outer isophotes, and
isophotal  
radii. Also listed, for 20 objects, is the $B$-band $D_{25}$ diameter
from the ESO-LV images, re-calibrated as explained in Sect. 3.

Tables 1, 2a, 2b, and 3 are available from the CDS, Strasbourg, via anonymous 
ftp to 130.79.128.5.
   \subsection{Surface photometry compared to other authors}
   \begin{enumerate}
   \item Brightness accuracy: from the catalog of Gezari et al. (1993), we 
collected the $K$-band aperture-photometry data available in the recent 
literature for six galaxies belonging to our sample (an additional object,
NGC 7172, was eliminated due to obvious inconsistencies between the data from
different sources and even within a single source). We performed simulated
aperture photometry on our images, and Table 4 presents the results compared
to previously published data, converted to $K'$. We found a mean difference of 
$0.038 \pm 0.073$, a reasonable agreement. 
For NGC 6810, we found an excellent agreement with the $H$
and $K$ (corrected to $K'$) 
measurements of Griersmith et al. (1982: hereafter GHJ82), but we noticed a 
systematic discrepancy of 0.12 mag in $J$ between their values and ours.
%
    \begin{table}
    \caption{Comparison with published aperture photometry}
    \begin{flushleft}
    \begin{tabular}{lrllrr}
    \hline
object & $A$ & $K'$ & Ref & $K'_{\rm mes}$ & residual\\
    \hline
(1)    & (2) & (3)  & (4) & (5) & (6)       \\
    \hline
IC 5267  & 21.8  &  9.16   & GHJ82 &  9.03 & $-0.13$\\
IC 5267  & 33.0  &  8.78   & GHJ82 &  8.71 & $-0.06$\\
IC 5267  & 56.0  &  8.38   & GHJ82 &  8.33 & $-0.05$\\
IC 5271  & 18.1  & 10.14   & GHJ82 &  9.96 & $-0.18$\\
IC 5271  & 30.1  &  9.48   & GHJ82 &  9.35 & $-0.13$\\
NGC 6810 & 12.0  &  9.18   & GHJ82 &  9.18 & $ 0.00$\\
NGC 6810 & 18.1  &  8.76   & GHJ82 &  8.76 & $ 0.00$\\
NGC 6810 & 30.1  &  8.36   & GHJ82 &  8.38 & $ 0.02$\\
NGC 7537 & 44.5  & 10.43   & BRS84 & 10.35 & $-0.08$\\
NGC 7541 & 44.5  &  9.26   & BRS84 &  9.24 & $-0.02$\\
NGC 7541 & 54.7  &  9.05   & BRS84 &  9.07 & $ 0.02$\\
NGC 7721 &  7.2  & 12.06   & DEV89 & 12.15 & $ 0.09$\\
NGC 7721 &  9.3  & 11.76   & DEV89 & 11.78 & $ 0.02$\\
    \hline
    \end{tabular}
    \end{flushleft}
{\it Columns\/}:

(1) galaxy identification\\
(2) $A$: aperture diameter (in arcsec)\\
(3) $K'$: magnitude within $A$\\
(4) Ref: reference for Col. (3); BRS84: Bothun et al. (1984); DEV89:
Devereux (1989); GHJ82: Griersmith et al. (1982)\\
(5) $K'_{\rm mes}$: our measured value for simulated aperture photometry
within  
$A$\\
(6) residual: $K'_{\rm mes} - K'$
    \end{table}
%
We also compared our profiles, for
three galaxies in common (NGC 7537, NGC 7541, and NGC 7606),
with those of Terndrup et al. (1994), who also used ellipse fitting to the
isophotes. We applied a  
suggested correction of 0.65 mag to their data (Terndrup, private 
communication), and another $K$-to-$K'$ correction of 0.05 mag; we found a 
good agreement (Fig. 2), except for the innermost regions where resolution 
effects may be present, and for the intermediate region of NGC 7541 (between
$\simeq 20''$ and $\simeq 30''$ from the center).
Globally, if a systematic offset exists, it does not exceed 0.1 mag. 
%
   \begin{figure}
   \epsfxsize=8.8cm
   \epsffile[77 240 474 446]{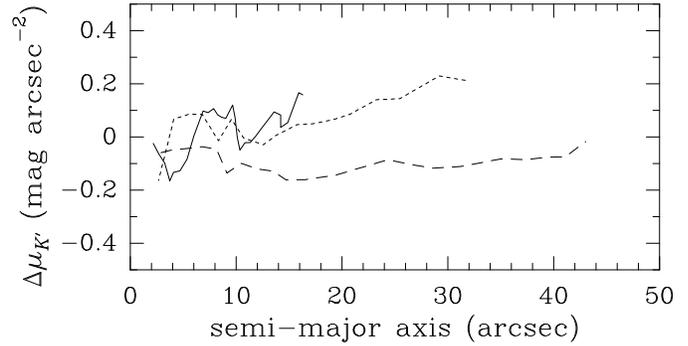}
   \caption{Comparison of our $K'$ photometric profiles with those of
Terndrup et al. (1994), for the three galaxies in common. The magnitude 
difference
$\Delta\mu_{K'}=\mu_{\rm (ours)}-\mu_{\rm (theirs)}$ is plotted as a function
of the semi-major axis of the isophotes. 
{\em Solid line}\/: NGC 7537;
{\em Dotted line}\/: NGC 7541;
{\em Dashed line}\/: NGC 7606. Data at less than $2''$ from the center are not
shown}
   \end{figure}
%
   \item Determination of the major-axis position angle: we compared our
measured values with those listed in the {\it LEDA\/} database, and we found
for four objects (NGC 5604, NGC 5937, NGC 7721, and MCG -1-39-05), a
significant 
discrepancy with the cited source (Corwin, to be published), in the 
sense that $PA_{\rm ours} + PA_{LEDA} \simeq 180^\circ$. 
With these exceptions,
the rms difference in the $PA$ values is \, $\simeq 5 \degr$ (we note that 
this figure includes the effect due to the difference in morphology between
the NIR and visible domains). 
   \end{enumerate}

   \subsection{Morphology}
With $M_B \la -19$, our galaxies are intrinsically bright; the sample
is limited but, not surprisingly, the global characteristics statistically
confirm what has recently been published on the
NIR morphology (see, e.g., Block \& Wainscoat 1991; Rix \& Rieke 1993;
Block et al. 1994; de Jong 1995; Rauscher 1995). Our main comments are the 
following: 
   \begin{itemize}
   \item we found a conspicuous bar in three objects so far classified 
unbarred (ESO 323-38, MCG 1-39-05 and MCG 2-38-22), and there are 
five suspected other cases (IC 4821, IC 5271, NGC 5119, NGC 5604 and NGC 6063);
the bars are much more frequent than can be estimated in the visible. But we
also noted one opposite case (NGC 7541), where the barred structure seen in
the visible is only an artifact of the obscuring pattern.
   \item three galaxies exhibit a central bright, flattened feature exactly
aligned along the major axis, which is likely to be disky (ESO 320-26, NGC
6810, 
and NGC 7172); two other cases are more ambiguous (ESO 380-06 and NGC 7083). 
it is not clear whether this is
only the inward prolongation of the main disk (made visible by the low 
extinction) or an independent inner structure. This question demands a
thorough photometric analysis which is beyond the scope of the present paper;
for one object, ESO 320-26, H\'eraudeau et al. (1995b: hereafter HSM95b) 
present evidence for an independent structure.
   \item the spiral pattern, sometimes blurred on images in the visible bands,
is more contrasted, and its general aspect fits into the framework defined
by Block et al. (1994). A few galaxies, however, exhibit particularly
thin arms, which may deserve a closer attention (NGC 4219, NGC 7083, NGC 7541,
NGC 7606, and MCG 2-38-22).
   \item photometric profiles often show the distinctive characteristic known 
as ``Type II'' (Freeman 1970): this is the case for at least 11 out of the 31 
galaxies in our sample. It is sometimes more pronounced in the visible 
than in the NIR (effect of the young population?), but it is sometimes the 
opposite (effect of dust?).
   \end{itemize}
   \subsection{Color-color diagrams}
%
   \begin{figure}
   \epsfysize=15cm
   \epsffile[0 113 340 700]{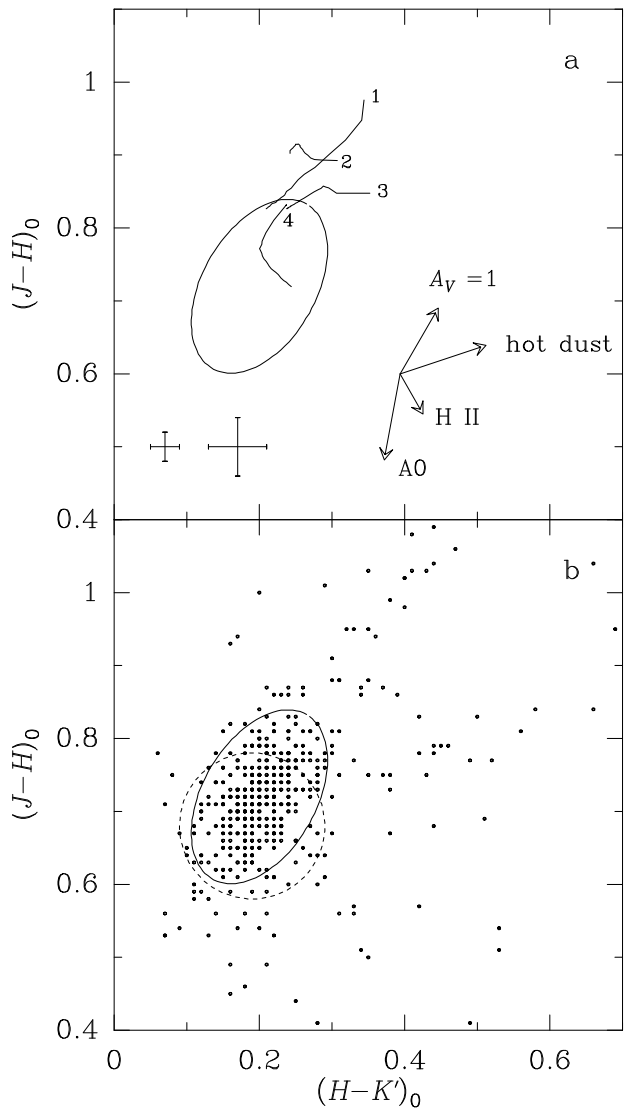}
   \caption{Color-color diagrams. The indices are corrected for Galactic
extinction and redshift.
{\bf a} data for our four galaxies; {\it solid
curves\/}: loci of color indices integrated within isophotes, the numbers
are close to the innermost points and identify the objects (1, NGC 4219;
2, NGC 6788; 3, NGC 6810; 4, ESO 320-G 026); {\it error bars\/}: mean
uncertainty on indices, at 20\arcsec \, and 40\arcsec \, from the center;
{\it arrows\/}: effects of Galactic-law extinction, and 10\% of flux from:
hot (1000 \degr K) dust, thermal bremsstrahlung (H {\sc ii}), and 
A0 stars (10\,000 \degr K blackbody);
{\it ellipse\/}: mean area of non-AGN spirals.
{\bf b} data from the literature; {\it points\/}:
aperture photometry for non-AGN spirals from de Vaucouleurs \& Longo (1988);
{\it solid ellipse\/}: mean area; {\it dashed circle\/}: mean area
corresponding to elliptical galaxies}
   \end{figure}
%
For the four objects observed in three passbands, the angular resolution 
was sufficient to allow an examination of the color maps, and we wanted to
use the color-color diagrams to get an insight on the obscuration by dust.
We calculated the mean $(J-H)_0$ and $(H-K')_0$ indices integrated within
ellipses fitted to the isophotes; the subscript 0 indicates values corrected
for Galactic extinction (Rieke \& Lebofsky 1985, and references therein),
and redshift ($K$-correction: Longmore \& Sharples 1982). 
The parameters of the ellipses were the same for the three colors; they were 
determined on the $K'$-band images, beginning at a semi-major axis of 
2\arcsec, with a step of 2\arcsec. Our results are displayed in Fig. 3a.

We presented in Fig. 3b, for comparison, the diagrams corresponding to:
\begin{itemize}
\item aperture-photometry of elliptical galaxies from the compilation of
de Vaucouleurs \& Longo (1988); this represents 258 measurements at different
apertures on 96 objects (the following three references gather $\simeq 70\%$
of these data: Aaronson 1977; GHJ82; Glass 1984). The same corrections as above 
were applied, together with a $K$-to-$K'$ correction.
\item aperture-photometry of 
spirals from the same compilation, selected for being non-AGN
objects according to V\'eron-Cetty \& V\'eron (1993): 435 measurements on 
191 objects. A well-known characteristic, already noted by, e.g., GHJ82, 
Glass (1984), and Carico et al. (1984), is clearly present: the clustering
of most of the data in a region of the graph close to that occupied by the 
elliptical galaxies; we may be able to add a point: that the mean shift
from the elliptical to spiral barycenters corresponds to $A_V \la 1$. 
Besides the accumulation of data in this region, the spirals also show
a tail extending to the redder end of the diagram, and which corresponds
to the inner regions of $\simeq 20\%$ of the galaxies (GHJ82; Glass 1984). 
Many galaxies of the {\it IRAS\/} minisurvey (Soifer et al. 1984; 
Rowan-Robinson et al. 1984) populate this tail (Carico et al. 1986; Moorwood
et al. 1986, 1987); Seyfert galaxies occupy the lower envelope of the tail 
(GHJ82, and references therein; Kotilainen et al. 1992). 
\end{itemize}

Compared to the data of Fig. 3b, our results for the four galaxies lead to
the following comments: 
   \begin{itemize}
   \item The global extent of the color-color loci, although not contained
in the mean ellipse, is nevertheless close to it, and within a region
populated by many other spirals. The four galaxies do not belong to the
sample of the minisurvey, but they have $L_{\rm FIR} / L_B$ ratios within its
range (although close to the lower limit), and they lie within the
characteristic tail. NGC 4219, which is ranked second in 
$L_{\rm FIR} / L_B$ ratio, reaches the redder color indices, along both axes. 
NGC 6810, with the highest ratio, is found in the literature with
either Seyfert-2 or starburst-nucleus classifications, and is well within the
region of the diagram occupied by such objects.
   \item We may note (naively), that various physical processes and population 
effects are available for explaining the observed tracks in the 
color-color diagrams; there has been some evidence, however, that for normal
and 
``nearly-normal'' spirals, the $JHK$ diagrams can, at least crudely, be
interpreted in terms of reddening due to dust (GHJ82; Moorwood et al. 1986,
1987). 
Then, for at least three of our galaxies, the global extinction in 
the inner region can be estimated to roughly $A_V \la 1$ or 2 mags.
   \item A more detailed analysis of the absorption patterns (dust lanes and
patches), e.g. by using local color-color diagrams and a wider range of
passbands, is beyond the scope of this work, and will be presented later 
(HSM95b).

   \end{itemize}

   \subsection{Discussion}
We have presented NIR surface photometry for a sample of 31 spiral galaxies,
and comparison with (a few) other sources does not show significant systematic 
errors. Tables containing $K'$-band profiles (plus $J$ and $H$ for four
selected objects), isophote parameters, $B$ and $R$ profiles (some of them
corrected) from ESO-LV, and $I$ profiles (mostly from Mathewson et al. 1992)
are available in electronic form.

Morphological analysis confirms several global trends revealed by
recent studies, concerning the aspect of the spiral pattern and the bar
phenomenon. A brief look at color indices in the central region of the
four objects with large angular dimension suggests extinction effects at a
level not unexpected; these regions turned out to be rich in features like
dust lanes, dust patches, crescent-shaped structures.

Our data are aimed at preparing the following applications: 
   \begin{enumerate} 
   \item The surface brightness will be analyzed in term of stellar components.
Following Simien \& H\'eraudeau (1994) and H\'eraudeau \& Simien (1995: 
hereafter HS95), a
decomposition into bulge, axisymmetric disk and spiral-arm contributions will
be performed (H\'eraudeau et al. 1995a); the hypothesis that the ``Freeman
type II'' profiles are associated to the spiral pattern (HS95; de Jong 1995)
will 
be tested. And the inner bright disks, as observed in several members of our
sample, will deserve particular attention. Our results will be compared to
recent other works, e.g., Peletier et al. (1994), and de Jong (1995) for
face-on galaxies.
   \item The mass models resulting from the photometric decomposition will be
used to generate rotation curves to be compared to $\rm H\alpha$ and H
{\sc i} measurements (H\'eraudeau et al. 1995a). Taking advantage of 
a) the better mass dependence of the NIR surface brightness in the central and
intermediate regions (Rix \& Rieke 1993), and, b) the farther-reaching 
$B$, $R$, or $I$ data in the outer regions, should allow a fair estimate of 
the ``visible'' mass; constraints on the amount of dark matter will then
result.  
   \item Multi-band data, from $B$ to $K'$, will be used to investigate
extinction effects in more detail, and central features will be studied
individually (HSM95b).
   \end{enumerate} 
   \acknowledgements{We thank R.F. Peletier and E.A. Valentijn
for their valuable advice on infrared techniques prior to our observing run. 
We are indebted to R.F. Peletier who, as the referee, made numerous comments
that helped to improve the manuscript. We are also 
pleased to thank D. Mathewson and V. Ford for providing us with their 
data in digital form, the ESO-ST/ECF Archive Service for extracting images
from the ESO-LV database, and Ph. Prugniel and E. P\'econtal for helpful 
discussions.
This research was sponsored in part by a grant (to G.A.M.) from the GdR
Cosmologie (CNRS).
The {\it LEDA\/} extragalactic database is operated by G. Paturel and the Lyon
and Meudon Observatories.}

\vspace{0.5cm}
{\noindent \bf Appendix A.}

\vspace{0.1cm}
\noindent Comments on individual objects:

   \begin{enumerate}
   \item {\bf ESO 320-26}: pointed innermost isophotes ($e_4 \simeq 
0.04$), providing evidence for a bright central disk; this inner structure
appears too bright to be the inward prolongation of the main disk (HSM95b).
Boxy bulge structure? ``Type II'' photometric profile.
   \item {\bf ESO 380-06}: boxy bulge. Possible inner disk. 
   \item {\bf ESO 267-37 / NGC 4219}: outer warp visible on the $B$ 
image; two-armed inner spiral pattern, with emergence of arms at 
$\simeq 4\arcsec$ from the nucleus. Thin main spiral arms. Evidence for a 
compact bulge surrounded
by a crescent-shaped structure of diameter $\simeq 7$\arcsec\ parallel to
the major axis. ``Type II'' profile.
   \item {\bf ESO 322-28}: many foreground stars superposed. Bright, 
compact bulge. ``Type II'' profile in $B$ and $R$, not in $K'$.
   \item {\bf ESO 323-38}: its structure appears rather patchy for a 
$K'$-band image. Evidence of a bar of diameter $\simeq 4\arcsec$ aligned
with outer disk. ``Type II'' profile.
   \item {\bf NGC 5119}: spiral arms almost parallel to the major axis. 
Possibly barred. May have been classified too late at T=5. ``Type II'' profile.
   \item {\bf NGC 5604}: may be slightly barred, although classified SAa. 
And a later type may be more appropriate. ``Type II'' profile.
   \item {\bf MCG 2-38-22}: thin bar, with thin arms starting from its
extremities and winding for almost 180\degr. Very compact bulge.
   \item {\bf MCG 1-39-05}: thin, elongated bar.
   \item {\bf NGC 5937}: although a short $m=2$ pattern dominates 
the spiral structure, a third arm is clearly apparent; overall patchy aspect,
unusually similar to a visible-band image.
   \item {\bf NGC 6063}: poor S/N ratio, but a ``grand-design'' spiral 
pattern is detectable. Compact bulge. Possible inner bar.
   \item {\bf ESO 139-11}: observed during critical weather conditions, 
and a cloud is visible on the frame, although not on the object itself. Compact
bulge. Too small for detailed morphological analysis. ``Type II'' profile.
   \item {\bf ESO 140-24 / IC 4717}: the PSF turned out to be asymmetric. 
Highly inclined. Suspected ringed structure (barely visible) or, conversely,
spiral pattern mimicking a ring. ``Type II'' profile.
   \item {\bf ESO 184-17 / IC 4821}: many superposed stars. The elongated
central condensation has not exactly the same orientation as the major axis,
and may be a bar; this is not visible on the $B$ and $R$ images.
   \item {\bf ESO 184-27 / IC 4826}: only the central region is available 
on our $K'$ image. No apparent structure.
   \item {\bf ESO 184-67 / NGC 6788}: emergence of embryonic spiral arms 
at $r \simeq 5$ \arcsec; inside, the bulge position angle is very close to  
that of the outer disk, suggesting an oblate geometry. No evident continuity 
of the spiral arms.
   \item {\bf ESO 142-35 / NGC 6810}: highly inclined. In $B$, 
large-scale 
dust lane parallel to the major axis, and overall patchy appearance. In the 
NIR, complex central structure, with (i) pointed innermost isophotes, giving 
evidence for a bright disk here, (ii) a crescent-like bright feature at 
$\simeq 5\arcsec$ on the far side (barely visible in $B$), which turned up
to be the unobscured half of the innermost bulge region (HSM95b),
(iii) the emergence of two spiral arms, and (iv) a large bulge, 
whose outer regions distort the outermost disk isophotes (this 
far-reaching bulge is markedly different from the one visible in $B$ on the 
ESO-LV image; see, e.g., Simien et al. 1993).
   \item {\bf ESO 073-22 / IC 4929}: imperfect flatfielding. Highly 
inclined object, strongly asymmetric along the major axis, a characteristic 
also conspicuous on the ESO-LV images. Very dusty in $B$ and $R$; in $K'$, 
two symmetric inner arms are clearly visible. ``Type II'' profile.
   \item {\bf ESO 233-41 / NGC 6870}: amorphous disk structure. The 
bright, compact inner bulge is significantly dimmed on the $B$ image.
   \item {\bf ESO 399-25}: imperfect flatfielding; nevertheless, the 
profile seems reliable down to $\mu_{K'} \simeq 19.5$. No apparent structure.
   \item {\bf ESO 287-13}: observed with imperfect focusing. Shows 
marginal evidence of incipient spiral structure starting at 
$r \simeq 5\arcsec$. Indication of ``type II'' profile.
   \item {\bf ESO 107-36 / NGC 7083}: thin spiral structure; on our image,
it is difficult to make a choice between the following two interpretations:
either a two-armed structure winding for more than 180\degr, or a shorter
two-armed structure branching into a four-armed one at $\simeq 10\arcsec$ 
from the center. Bright, compact bulge; slightly pointed innermost isophotes 
($e_4 \simeq 0.03$), providing marginal evidence for a bright central disk.
   \item {\bf ESO 466-38 / NGC 7172}: nearly edge-on galaxy, with a strong
absorbing 
lane in $B$ parallel to the major axis (extinction $\Delta \mu_B \simeq 2$). 
In $K'$, the lane is barely visible; bright central disk (inner isophotes with
$e_4 \simeq 0.07$).
   \item {\bf NGC 7328}: asymmetric arms; faint bulge.
   \item {\bf ESO 290-29 / IC 5267}: dusty, low-contrast spiral pattern in
$B$, invisible in $K'$ (more akin an elliptical in this band).
   \item {\bf ESO 406-34 / IC 5271}: featureless in $B$. In $K'$, at
$r \simeq 7\arcsec$, marginal evidence for either the emergence of spiral arms
or the presence of a bar. The disk is not very smooth. Photometric profile 
down to $\mu_{K'} = 22$. ``Type II'' profile.
   \item {\bf NGC 7537}: diffuse; very faint bulge (classified too late?).
   \item {\bf NGC 7541}: nearly edge-on, with a strong absorbing lane in 
$B$. Classified SB, but we see no evidence for a bar. Strong
asymmetry along the two semi-major axes; thin asymmetric arms; compact bulge. 
``Type II'' profile?
   \item {\bf NGC 7606}: multiple, tightly-wound thin spiral arms. Small bulge 
for a type T=3. ``Type II'' profile.
   \item {\bf ESO 605-07 / IC 5321}: the innermost profile only could be 
accurately measured (out to $r \simeq 15 \arcsec$).
   \item {\bf NGC 7721}: asymmetric arms. Highly flattened bulge, or bar 
oriented along the major axis? ``Type II'' profile.
   \end{enumerate}

\vspace{0.5cm}
{\noindent \bf Appendix B.}
\vspace{0.1cm}

\noindent Table 5 presents the brightness of the faintest isophotes displayed 
in the images of Fig. 4, in mag arcsec$^{-2}$.

    \begin{table}
    \caption{Faintest-level isophotes in Fig. 4}
    \begin{flushleft}
        \begin{tabular}{rllcccc}
        \hline
 \# & \multicolumn{2}{c}{object}& $B_f$ & $J_f$ & $H_f$ & $K'_f$ \\
        \hline
 1 & E 320-26    &                 & 22.0      &19.0   & 18.5  & 18.0 \\
 2 & E 380-06    &                 & 22.0      &       &       & 17.5 \\
 3 & E 267-37    &  N 4219         & 23.0      & 19.5  & 18.5  & 18.0  \\
 4 & E 322-28    &                 & 22.5      &       &       & 18.5 \\
 5 & E 323-38    &                 & 23.0      &       &       & 18.5 \\
 6 &             &  N 5119         &           &       &       & 18.5 \\
 7 &             &  N 5604         &           &       &       & 18.5 \\
 8 &             & M -2-38-22      &           &       &       & 18.5 \\
 9 &             & M -1-39-05      &           &       &       & 18.0 \\
10 &             & N 5937          &           &       &       & 18.5 \\
11 &             & N 6063          &           &       &       & 19.0 \\
12 & E 139-11    &                 & 22.5      &       &       & 18.5 \\
13 & E 140-24    & I 4717          &   23.0    &       &       & 19.0 \\
14 & E 184-17    & I 4821          &   22.0    &       &       & 18.5 \\
15 & E 184-27    & I 4826          &   22.5    &       &       & 19.0 \\
16 & E 184-67    & N 6788          & 22.5      & 19.5  & 18.5  & 18.5  \\
17 & E 142-35    & N 6810          & 22.5      & 20.0  & 19.0  & 18.5  \\
18 & E 073-22    & I 4929          &   22.0    &       &       & 19.0 \\
19 & E 233-41    & N 6870          &   21.5    &       &       & 18.5 \\
20 & E 399-25    &                 & 23.0      &       &       & 19.0 \\
21 & E 287-13    &                 & 22.0      &       &       & 18.0 \\
22 & E 107-36    & N 7083          &   21.5    &       &       & 18.0 \\
23 & E 466-38    & N 7172          &   22.5    &       &       & 18.0 \\
24 &             & N 7328          &           &       &       & 18.0 \\
25 & E 290-29    & I 5267          &  21.5     &       &       & 17.5 \\
26 & E 406-34    & I 5271          &   22.0    &       &       & 18.0 \\
27 &             & N 7537          &           &       &       & 18.5 \\
28 &             & N 7541          &    22.0   &       &       & 18.0 \\
29 &             & N 7606          &           &       &       & 17.5 \\
30 & E 605-07    & I 5321          &  22.0     &       &       & 18.5 \\
31 &             & N 7721          &           &       &       & 18.0 \\
\hline
\end {tabular}
\end{flushleft}
\end{table}
   
   \begin{figure*}
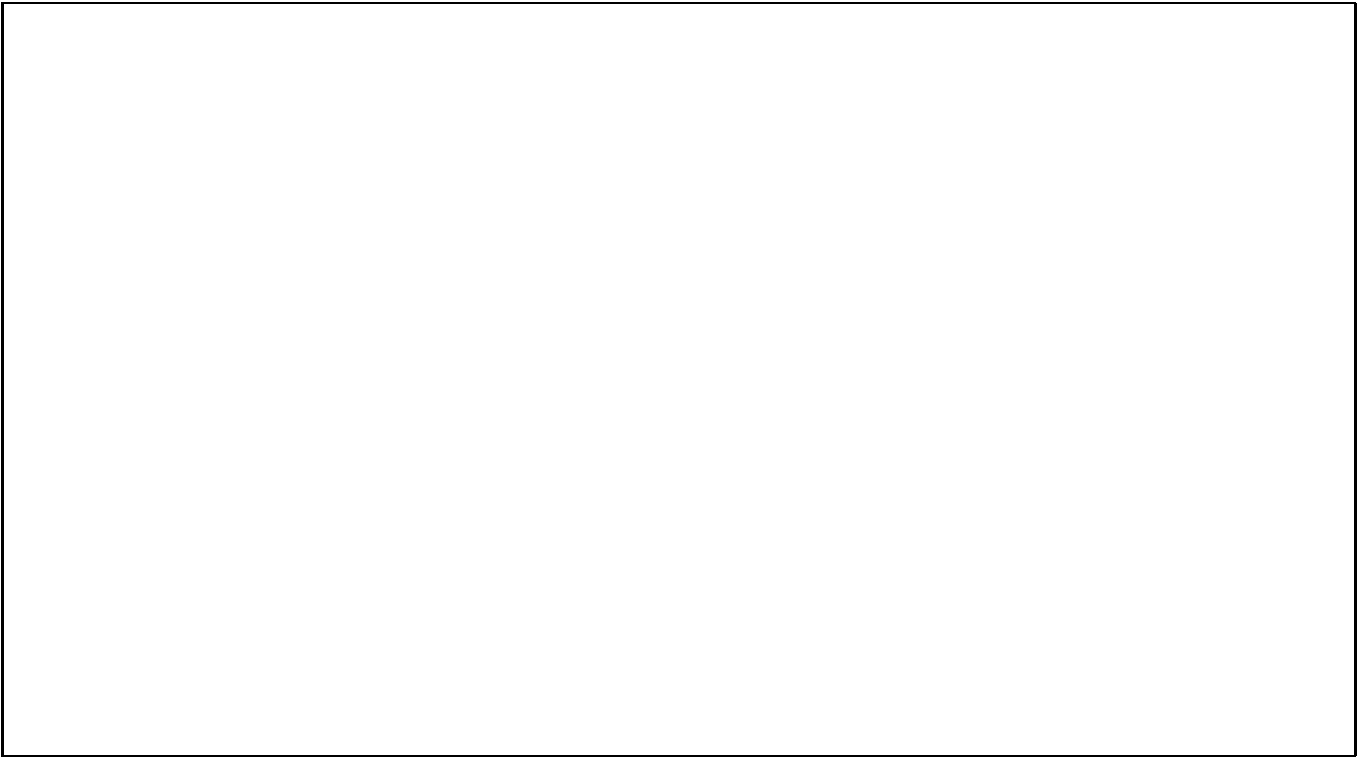

   \picplace{10cm}
\caption{Individual galaxies.
{\bf Greyscales}: $B$ image (from ESO-LV, when available, with pixel
size $1.35\arcsec$; from HS96 for galaxy \#28, NGC 7541, with pixel 
size $0.78\arcsec$) and $K'$
image; for four objects, $J$ and $H$ images are also given. 
North is on top and East is to the left. Contours and greylevels
are spaced at integer and half-integer values of surface magnitude.
Faintest isophotes are listed in Table 5 and chosen to produce high
signal-to-noise contours in all bands and similar extension.
{\bf Surface brightness profiles}: $\mu_{K'}$ ({\sl full circles\/}), and when
available,
$\mu_B$ ({\sl open triangles\/}, from ESO-LV),
$\mu_V$ ({\sl open squares\/}, from HS96),
$\mu_R$ ({\sl open circles\/}, from ESO-LV),
$\mu_I$ ({\sl crosses\/}, from Mathewson et al. 1992, except for galaxy \#28,
which is from HS96),
$\mu_J$ ({\sl full squares\/}), $\mu_H$ ({\sl full triangles\/}).
Error bars are only given for the NIR profiles.
{\bf Color profiles}: when available,
$B-K'$ ({\sl full circles\/}),
$V-K'$ ({\sl open squares\/}),
$R-K'$ ({\sl open circles\/}),
$I-K'$ ({\sl crosses\/}).
{\bf Isophote-parameter profiles} (for $K'$ isophotes):
ellipticity $\epsilon$, position angle $PA$, $e_4$ (when available), and 
also $e_6$ for the four large galaxies with $JHK'$ data.
Ellipticity and position angle from {\it LEDA\/} are shown as {\sl thin
horizontal lines\/} on right-hand side of plot
}
   \end{figure*}
   \end{document}